\documentclass[12pt]{article}
\usepackage[dvips]{graphics}

\catcode`@=11
\def\@citex[#1]#2{\if@filesw\immediate\write\@auxout{\string\citation{#2}}\fi
  \def\@citea{}\@cite{\@for\@citeb:=#2\do
    {\@citea\def\@citea{,\penalty\@m}\@ifundefined
      {b@\@citeb}{{\bf ?}\@warning
       {Citation `\@citeb' on page \thepage \space undefined}}%
\hbox{\csname b@\@citeb\endcsname}}}{#1}}

\def\citer{\@ifnextchar
[{\@tempswatrue\@citexr}{\@tempswafalse\@citexr[]}}
\def\@citexr[#1]#2{\if@filesw\immediate\write\@auxout{\string\citation{#2}}\fi
  \def\@citea{}\@cite{\@for\@citeb:=#2\do
    {\@citea\def\@citea{--\penalty\@m}\@ifundefined
       {b@\@citeb}{{\bf ?}\@warning
       {Citation `\@citeb' on page \thepage \space undefined}}%
\hbox{\csname b@\@citeb\endcsname}}}{#1}}
\catcode`@=12

\newcommand{\be}{\begin{equation}}
\newcommand{\ee}{\end{equation}}
\newcommand{\bea}{\begin{eqnarray}}
\newcommand{\eea}{\end{eqnarray}}
\newcommand{\ba}{\begin{array}}
\newcommand{\ea}{\end{array}}

\newcommand{\plb}[2]{{\em Phys. Lett.}              {\bf #1B}, #2 }
\newcommand{\npb}[2]{{\em Nucl. Phys.}              {\bf B#1}, #2 }

\newcommand{\pr }[2]{{\em Phys. Rep.}               {\bf  #1}, #2 }
\newcommand{\pra}[2]{{\em Phys. Rev.}               {\bf A#1}, #2 }
\newcommand{\prd}[2]{{\em Phys. Rev.}               {\bf D#1}, #2 }
\newcommand{\prl}[2]{{\em Phys. Rev. Lett.}         {\bf  #1}, #2 }
\newcommand{\zpc}[2]{{\em Z. Phys.}                 {\bf C#1}, #2 }
\newcommand{\sci}[2]{{\em Science}                  {\bf  #1}, #2 }
\newcommand{\jpb}[2]{{\em J. Phys.}                 {\bf B#1}, #2 }

\newcommand{\mpl}[2]{{\em Mod. Phys. Lett.}         {\bf A#1}, #2 }
\newcommand{\rmp}[2]{{\em Rev. Mod. Phys.}          {\bf  #1}, #2 }

\newcommand{\epj}[2]{{\em Eur. Phys. J.}            {\bf C#1}, #2 }
\newcommand{\con}[2]{                               {\bf  #1}, #2 }
\newcommand{\etal}{{\em et al.}}
\newcommand{\ibid}{{\em ibid.}}
\newcommand{\col}{Collaboration}

\renewcommand{\LARGE}{\large}

\begin{document}

\hfill \\ \vspace{10pt} 
\hfill {\normalsize UPR--839--T} \\
\vspace{20pt}
\centerline{\LARGE \bf Constraints on Extended Neutral Gauge Structures}
\vspace{15pt}
\centerline{\large Jens Erler and Paul Langacker}
\centerline{
\it Department of Physics and Astronomy, University of Pennsylvania,}
\centerline{\it Philadelphia, PA 19104-6396}
\centerline{\it E-mail: erler@langacker.hep.upenn.edu; 
                          pgl@langacker.hep.upenn.edu}

\begin{abstract}
Indirect precision data are used to constrain the masses of possible extra
$Z^\prime$ bosons and their mixings with the ordinary $Z$. We study a variety 
of $Z^\prime$ bosons as they appear in $E_6$ and left-right unification models,
the sequential $Z$ boson, and the example of an additional $U(1)$ in a concrete
model from heterotic string theory. In all cases the mixings are severely 
constrained ($\sin\theta < 0.01$). The lower mass limits are generally of the
order of several hundred GeV and competitive with collider bounds. The 
exception is the $Z_\psi$ boson, whose vector couplings vanish and whose limits
are weaker. The results change little when the $\rho$ parameter is allowed, 
which corresponds to a completely arbitrary Higgs sector. On the other hand, 
in specific models with minimal Higgs structures the limits are generally 
pushed into the TeV region. 
\end{abstract}

The possibility of additional neutral gauge bosons, $Z^\prime$s, is among the 
best motivated types of physics beyond the Standard Model (SM).
They are predicted by most unifying theories, such as Grand Unified Theories 
(GUTs), left-right unification, superstring theories and their strong coupling 
generalizations. In many cases their masses remain unpredicted and may or
may not be of the electroweak scale. In the context of superstring models,
however, which are much more constrained than purely field theoretical models,
they are often predicted to arise at the electroweak scale, as we will discuss 
below. In this paper we consider six different types of $Z^\prime$ bosons:
\begin{enumerate}
  \item The $Z_\chi$ boson is defined by $SO(10) \rightarrow SU(5) \times 
        U(1)_\chi$. This boson is also the unique solution to the conditions of
        (i) family universality, 
        (ii) no extra matter other than the right-handed neutrino,
        (iii) absence of gauge and mixed gauge/gravitational anomalies, and
        (iv) orthogonality to the hypercharge generator. In the context of
        a minimal $SO(10)$ GUT, conditions (i) and (ii) are satisfied by 
        assumption, while (iii) and (iv) are automatic. Relaxing condition 
        (iv) allows other solutions (including the $Z_{LR}$ below) which 
        differ from the $Z_\chi$ by a shift proportional to the third component
        of the right-handed isospin generator. 
  \item The $Z_\psi$ boson is defined by $E_6 \rightarrow SO(10) \times 
        U(1)_\psi$. It possesses only axial-vector couplings to the ordinary
        fermions. As a consequence it is the least constrained of our examples.
  \item The $Z_\eta$ boson is the linear combination 
        $\sqrt{3/8}\, Z_\chi - \sqrt{5/8}\, Z_\psi$. It occurs in Calabi-Yau 
        compactifications~\cite{Candelas85} of the heterotic 
        string~\cite{Gross85} if $E_6$ breaks directly to a rank~5 
        subgroup~\cite{Witten85} via the Hosotani mechanism~\cite{Hosotani83}.
  \item The $Z_{LR}$ boson occurs in left-right models with gauge group 
        $SU(3)_C \times SU(2)_L \times SU(2)_R \times U(1)_{B-L}\subset SO(10)$
        and is defined through the current 
        $J_{LR} = \sqrt{3/5}\, [\alpha J_{3R} - 1/(2\alpha) J_{B-L}]$. $J_{3R}$
        couples to the third component of $SU(2)_R$, $B$ and $L$ coincide with 
        baryon and lepton number for the ordinary fermions, 
        $\alpha = \sqrt{g_R^2/g_L^2 \cot^2 \theta_W - 1}$, where $g_{L,R}$ are 
        the $SU(2)_{L,R}$ gauge couplings, 
        and $\theta_W$ is the weak mixing angle.
  \item The {\em sequential\/} $Z_{SM}$ boson is defined to have the same 
        couplings to fermions as the SM $Z$ boson. Such a boson is not expected
        in the context of gauge theories unless it has different couplings to 
        exotic fermions than the ordinary $Z$. However, it serves as a useful
        reference case when comparing constraints from various sources. It 
        could also play the role of an excited state of the ordinary $Z$
        in models with extra dimensions at the weak scale. 
  \item Finally we consider a superstring motivated $Z_{string}$ boson 
        appearing in a specific model~\cite{Chaudhuri95} 
\parbox{100pt}{\vspace{2pt}
\begin{tabular}{|c|r|}
  \hline
  multiplet & $100\, Q^\prime$ \\ \hline
  $\left( \begin{array}{c} t \\ b \end{array} \right)_L$ & $- 71$ \\
  $t_R$                                                  & $+133$ \\
  $b_R$                                                  & $-136$ \\
  $\left( \begin{array}{c} u \\ d \end{array} \right)_L$,
  $\left( \begin{array}{c} c \\ s \end{array} \right)_L$ & $+ 68$ \\
  $u_R$, $c_R$                                           & $-  6$ \\
  $d_R$, $s_R$                                           & $+  3$ \\
  $\left( \begin{array}{c} \nu_\tau \\ \tau \end{array} \right)_L$ & $+ 74$ \\
  $\tau_R$                                                         & $-130$ \\
  $\left( \begin{array}{c} \nu_\mu  \\ \mu  \end{array} \right)_L$ & $- 65$ \\
  $\mu_R$                                                          & $+  9$ \\
  $\left( \begin{array}{c} \nu_e    \\ e    \end{array} \right)_L$ & $-204$ \\
  $e_R$                                                            & $+  9$ \\
  \hline
\end{tabular}} \hfill
\parbox{291pt}{\vspace{2pt} based on the free 
        fermionic string construction with real fermions. This model has been
        investigated in considerable detail~\cite{Cleaver98} with the goal of 
        understanding some of the characteristics of (weakly coupled) string 
        theories, and of contrasting them with the more conventional ideas such
        as GUTs. While this specific model itself is not realistic (for example
        it fails to produce an acceptable fermion mass spectrum) the predicted 
        $Z_{string}$ it contains is not ruled out. Its coupling strength is 
        predicted and so are its fermion couplings. It is particularly 
        interesting in that its couplings are family non-universal. While this 
        may induce problems with too large flavor changing neutral currents 
        through a violation of the GIM mechanism~\cite{Glashow70}, we will not 
        address this issue here. Another important observation is that such a 
        $Z_{string}$ can be naturally at the electroweak scale\cite{Cvetic96}. 
        The basic reasons are strong restrictions on the superpotential, and 
        that in a given model the sectors of supersymmetry (SUSY) breaking, 
        and its mediation are usually not arbitrary. In string models one} 

        \vspace{-2pt} also expects bilinear terms in the Higgs superpotential 
        to vanish at tree level (otherwise they should be of the order of the 
        Planck scale) and to be generated in the process of radiative symmetry
        breaking. This is linked to the top quark Yukawa coupling driven 
        symmetry breaking and typically involves extra Higgs singlets which 
        are predicted in many string models.
\end{enumerate}
In all cases, there is a relation between the mixing angle $\theta$
between the ordinary $Z$ and the extra $Z^\prime$ from the
diagonalization of the neutral vector boson mass matrix,
\be
\label{rel1}
   \tan^2 \theta = \frac{M_0^2 - M_Z^2}{{M_{Z^\prime}}^2 - M_0^2}
                 = \frac{1 - \rho_0/\rho_1}{\rho_0/\rho_2 - 1},
\ee
where $M_Z$ and $M_{Z^\prime}$ are the physical boson masses, and $M_0$ is the
mass of the ordinary $Z$ in the absence of mixing. The second equality in
relation~(\ref{rel1}) uses the neutral and charged boson mass interdependence
which reads at tree level,
\be
   M_\alpha = {M_W\over \sqrt{\rho_\alpha} \cos\theta_W}.
\ee
As in the case of the $SU(3)_C \times SU(2)_L \times U(1)_Y$ model,
\be
  \rho_0 = \frac{\sum_i (t_i^2 - t_{3i}^2 + t_i) |\langle \phi_i \rangle|^2}
                {\sum_i 2 t_{3i}^2               |\langle \phi_i \rangle|^2},
\label{rho}
\ee
where $t_i$ ($t_{3i}$) is (the third component of) the weak isospin of the
Higgs field $\phi_i$. $\rho_0 = 1$ if only $SU(2)$ Higgs doublets and singlets 
are present, in which case $M_0$ would be known independently. Nondegenerate 
$SU(2)$ multiplets of extra fermions and scalars affect the $W$ and $Z$ 
self-energies at the loop level, and therefore contribute to the $T$ 
parameter~\cite{Peskin90}. They can arise, for example, in $E_6$ models, and 
in their presence $\rho_0$ should be replaced by $\rho_0/(1 - \alpha(M_Z) T)$.

If the Higgs $U(1)^\prime$ quantum numbers are known, as well, 
there will be an extra constraint,
\be
  \theta = C {g_2\over g_1} {M_Z^2\over M_{Z^\prime}^2},
\ee
where $g_{1,2}$ are the $U(1)$ and $U(1)^\prime$ gauge couplings with
\be
  g_2 = \sqrt{5\over 3}\, \sin \theta_W \sqrt{\lambda}\, g_1.
\ee
$\lambda = 1$ if the GUT group breaks directly to
$SU(3) \times SU(2) \times U(1) \times U(1)^\prime$, while in general
$\lambda$ is still of ${\cal O}(1)$. We will quote our results assuming 
$\lambda = 1$, but our limits also apply to $\sqrt{\lambda}\, \sin\theta$ and
${M_{Z^\prime}\over \sqrt{\lambda}}$ for other values of $\lambda$
(we always assume $M_{Z^\prime} \gg M_Z$). Similar to $\rho_0$,
\be
  C = - \frac{\sum_i t_{3i} Q^\prime_i |\langle \phi_i \rangle|^2}
             {\sum_i t_{3i}^2          |\langle \phi_i \rangle|^2}
\ee
is another function of vacuum expectation values (VEVs), where $Q^\prime_i$ 
are the $U(1)^\prime$ charges. For minimal cases, the functions $C$ are given 
explicitly in Table~III of reference~\cite{Langacker92}. Similarly, for the 
fermion couplings of the various $Z^\prime$s we refer to Table~II of the same 
work. 

There is the possibility of an extra gauge invariant term, mixing the field 
strength tensors of the hypercharge and the new gauge bosons. 
We do not consider such a term here, since it is expected to be small in 
typical models\footnote{It was shown in Ref.~\cite{Babu98} that a relatively 
large kinetic mixing term can be generated at the loop level when Higgs 
doublets from a {\bf 78} representation of $E_6$ are employed. However, 
restriction to Higgs doublets from {\bf 27} and $\overline {\bf 27}$ 
representations yields much smaller effects~\cite{Langacker98}.}.
The phenomenology of gauge kinetic mixing has been 

\begin{table}[t]
\begin{center}
\begin{tabular}{|l|c|c|c|r|}
\hline Quantity & Group(s) & Value & Standard Model & pull \\ 
\hline
$M_Z$ \hspace{17pt}      [GeV]&     LEP     &$ 91.1867 \pm 0.0021 $&$ 91.1865 \pm 0.0021 $&$ 0.1$ \\
$\Gamma_Z$ \hspace{21pt} [GeV]&     LEP     &$  2.4939 \pm 0.0024 $&$  2.4957 \pm 0.0017 $&$-0.8$ \\
$\Gamma({\rm had})$\hspace{9pt}[GeV]&  LEP  &$  1.7423 \pm 0.0023 $&$  1.7424 \pm 0.0016 $&  ---  \\
$\Gamma({\rm inv})$\hspace{12pt}[MeV]& LEP  &$500.1    \pm 1.9    $&$501.6    \pm 0.2    $&  ---  \\
$\Gamma({\ell^+\ell^-})$ [MeV]&     LEP     &$ 83.90   \pm 0.10   $&$ 83.98   \pm 0.03   $&  ---  \\
$\sigma_{\rm had}$ \hspace{15pt}      [nb] &     LEP     &$ 41.491  \pm 0.058  $&$ 41.473  \pm 0.015  $&$ 0.3$ \\
$R_e$                         &     LEP     &$ 20.783  \pm 0.052  $&$ 20.748  \pm 0.019  $&$ 0.7$ \\
$R_\mu$                       &     LEP     &$ 20.789  \pm 0.034  $&$ 20.749  \pm 0.019  $&$ 1.2$ \\
$R_\tau$                      &     LEP     &$ 20.764  \pm 0.045  $&$ 20.794  \pm 0.019  $&$-0.7$ \\
$A_{FB} (e)$                  &     LEP     &$  0.0153 \pm 0.0025 $&$  0.0161 \pm 0.0003 $&$-0.3$ \\
$A_{FB} (\mu)$                &     LEP     &$  0.0164 \pm 0.0013 $&$                    $&$ 0.2$ \\
$A_{FB} (\tau)$               &     LEP     &$  0.0183 \pm 0.0017 $&$                    $&$ 1.3$ \\
\hline
$R_b$                         &  LEP + SLD  &$  0.21656\pm 0.00074$&$  0.2158 \pm 0.0002 $&$ 1.0$ \\
$R_c$                         &  LEP + SLD  &$  0.1735 \pm 0.0044 $&$  0.1723 \pm 0.0001 $&$ 0.3$ \\
$R_{s,d}/R_{(d+u+s)}$         &     OPAL    &$  0.371  \pm 0.023  $&$  0.3592 \pm 0.0001 $&$ 0.5$ \\
$A_{FB} (b)$                  &     LEP     &$  0.0990 \pm 0.0021 $&$  0.1028 \pm 0.0010 $&$-1.8$ \\
$A_{FB} (c)$                  &     LEP     &$  0.0709 \pm 0.0044 $&$  0.0734 \pm 0.0008 $&$-0.6$ \\
$A_{FB} (s)$                  &DELPHI + OPAL&$  0.101  \pm 0.015  $&$  0.1029 \pm 0.0010 $&$-0.1$ \\
$A_b$                         &     SLD     &$  0.867  \pm 0.035  $&$  0.9347 \pm 0.0001 $&$-1.9$ \\
$A_c$                         &     SLD     &$  0.647  \pm 0.040  $&$  0.6676 \pm 0.0006 $&$-0.5$ \\
$A_s$                         &     SLD     &$  0.82   \pm 0.12   $&$  0.9356 \pm 0.0001 $&$-1.0$ \\
\hline
$A_{LR}$ (hadrons)            &     SLD     &$  0.1510 \pm 0.0025 $&$  0.1466 \pm 0.0015 $&$ 1.8$ \\
$A_{LR}$ (leptons)            &     SLD     &$  0.1504 \pm 0.0072 $&$                    $&$ 0.5$ \\
$A_\mu$                       &     SLD     &$  0.120  \pm 0.019  $&$                    $&$-1.4$ \\
$A_\tau$                      &     SLD     &$  0.142  \pm 0.019  $&$                    $&$-0.2$ \\
$A_e (Q_{LR})$                &     SLD     &$  0.162  \pm 0.043  $&$                    $&$ 0.4$ \\
$A_\tau ({\cal P}_\tau)$      &     LEP     &$  0.1431 \pm 0.0045 $&$                    $&$-0.8$ \\
$A_e ({\cal P}_\tau)$         &     LEP     &$  0.1479 \pm 0.0051 $&$                    $&$ 0.3$ \\
$\bar{s}_\ell^2 (Q_{FB})$     &     LEP     &$  0.2321 \pm 0.0010 $&$  0.2316 \pm 0.0002 $&$ 0.5$ \\
\hline
\end{tabular}
\end{center}
\caption[]{$Z$ pole precision observables from LEP~\cite{Karlen98,Boudinov98} 
and the SLC~\citer{Baird98,Abe97A}. Shown are the experimental results, 
the SM predictions, and the pulls. The SM errors are from the uncertainties in 
$M_Z$, $\ln M_H$, $m_t$, $\alpha (M_Z)$, and $\alpha_s$. They have been treated
as Gaussian and their correlations have been taken into account. The first set 
of measurements is from the $Z$ line shape and leptonic forward-backward 
asymmetries, $A_{FB}(\ell) = 3/4 A_e A_\ell$. The hadronic, invisible, and 
leptonic decay widths are not independent of the total width, the hadronic peak
cross section, and the $R_\ell = \Gamma({\rm had})/\Gamma(\ell^+\ell^-)$, and 
are shown for illustration only. The second set represents the quark sector, 
where $R_q = \Gamma(q\bar{q})/\Gamma({\rm had})$, and 
$A_q = 4/3 A_{LR}^{FB} (q)$ is a function of the effective weak mixing angle
of quark $q$. The third set is a variety of polarization and forward-backward 
asymmetries sensitive to the leptonic weak mixing angle, where analogous 
definitions apply. For details see reference~\cite{Erler98}.}
\label{zpole}
\end{table}

\clearpage

\noindent
reviewed in 
reference~\cite{Babu98A}; constraints on leptophobic $Z^\prime$ bosons can be
found in reference~\cite{Umeda98}; reference~\cite{Leike98} are detailed 
reviews on extra neutral gauge bosons at colliders; and for related topics 
in neutral current physics see reference~\cite{Zeppenfeld98}. 

After the $Z^\prime$ properties have been specified, the new contributions to 
the precision observables can be computed using the formalism presented in 
references~\cite{Langacker92,Durkin86}. The new $Z^\prime$ boson is treated as 
a small perturbation to the SM relations. The most important effect is the 
modification of the $M_Z$--$M_W$--$\sin^2\theta_W$ interdependence via 
$Z$--$Z^\prime$ mixing. $Z$ pole observables are affected by the modification 
of the $Z$ couplings to fermions, which is another manifestation of the 
$Z^\prime$ admixture. This change in couplings is also relevant for the low 
energy observables from neutrino scattering and atomic parity violation (APV). 
For these there will be additional effects from $Z^\prime$ exchange, and
$Z$--$Z^\prime$ ($\gamma$--$Z^\prime$) interference. Such effects can be 
neglected at the $Z$ pole, but the interference terms are relevant for the 
cross section measurements at LEP 2. They give interesting constraints on 
$M_{Z^\prime}$~\cite{Renton99} practically independent of the mixing parameter
$\theta$. These constraints are complementary to the direct search limits at 
Fermilab~\cite{Abe97}, where additional assumptions about possible exotic decay
channels have to be specified. 

\begin{table}[bh]
\begin{center}
\begin{tabular}{|l|c|c|c|r|}
\hline Quantity & Group(s) & Value & Standard Model & pull \\ 
\hline
$m_t$\hspace{11pt}[GeV]&Tevatron &$ 173.8    \pm 5.0               $&$ 171.4    \pm 4.8    $&$ 0.5$\\
$M_W$ [GeV]    & Tevatron + UA2 &$  80.404  \pm 0.087             $&$  80.362  \pm 0.023  $&$ 0.5$\\
$M_W$ [GeV]    &      LEP       &$  80.37   \pm 0.09              $&$                     $&$ 0.1$\\
\hline
$R^-$          &     NuTeV      &$   0.2277 \pm 0.0021 \pm 0.0007 $&$   0.2297 \pm 0.0003 $&$-0.9$\\
$R^\nu$        &     CCFR       &$   0.5820 \pm 0.0027 \pm 0.0031 $&$   0.5827 \pm 0.0005 $&$-0.2$\\
$R^\nu$        &     CDHS       &$   0.3096 \pm 0.0033 \pm 0.0028 $&$   0.3089 \pm 0.0003 $&$ 0.2$\\
$R^\nu$        &     CHARM      &$   0.3021 \pm 0.0031 \pm 0.0026 $&$                     $&$-1.7$\\
$R^{\bar\nu}$  &     CDHS       &$   0.384  \pm 0.016  \pm 0.007  $&$   0.3859 \pm 0.0003 $&$-0.1$\\
$R^{\bar\nu}$  &     CHARM      &$   0.403  \pm 0.014  \pm 0.007  $&$                     $&$ 1.1$\\
$R^{\bar\nu}$  &     CDHS 1979  &$   0.365  \pm 0.015  \pm 0.007  $&$   0.3813 \pm 0.0003 $&$-1.0$\\
\hline
$g_V^{\nu e}$  &     CHARM II   &$  -0.035  \pm 0.017             $&$  -0.0395 \pm 0.0004 $&  --- \\
$g_V^{\nu e}$  &      all       &$  -0.041  \pm 0.015             $&$                     $&$-0.1$\\
$g_A^{\nu e}$  &     CHARM II   &$  -0.503  \pm 0.017             $&$  -0.5063 \pm 0.0002 $&  --- \\
$g_A^{\nu e}$  &      all       &$  -0.507  \pm 0.014             $&$                     $&$-0.1$\\
\hline
$Q_W({\rm Cs})$&     Boulder    &$ -72.41   \pm 0.25\pm 0.80      $&$ -73.10   \pm 0.04   $&$ 0.8$\\
$Q_W({\rm Tl})$&Oxford + Seattle&$-114.8    \pm 1.2 \pm 3.4       $&$-116.7    \pm 0.1    $&$ 0.5$\\
\hline
\end{tabular}
\end{center}
\caption[]{Non-$Z$ pole precision observables from
Fermilab~\citer{Dorigo98,McFarland98A}, CERN~\cite{Karlen98,Alitti92}, and 
elsewhere. The second error after the experimental value, where given, is 
theoretical. The SM errors are from the inputs as in Table~\ref{zpole}. The 
various quantities $R$ are cross section ratios from  $\nu$-hadron scattering, 
where the CHARM~\cite{Allaby87} results have been adjusted to 
CDHS~\cite{Blondel90} conditions, and can be directly compared. The $g_{V,A}$ 
are effective four-Fermi couplings from $\nu$-$e$ scattering~\cite{Vilain94}, 
and $Q_W$ denotes the weak charge appearing in 
APV~\citer{Wood97,Dzuba87}.}
\label{nonzpole}
\end{table}

In our analysis we use the data as of ICHEP~98 at Vancouver. It includes the 
very precise $Z$ pole measurements from LEP and the SLC, which are close to 
being finalized; the $W$ boson and top quark mass measurements, $M_W$ and 
$m_t$, from the Tevatron run~I, and further $M_W$ determinations from LEP 2; 
results from deep inelastic $\nu$-hadron scattering at CERN and Fermilab; 
$\nu$-electron scattering; and atomic parity violation. The low energy 
measurements in neutrino scattering and APV are very important in the presence
of new physics, and in particular, for the $Z^\prime$ bosons discussed here. 
They offer complementary information about $Z^\prime$ exchange and interference
effects, which are suppressed at the $Z$ pole. The $Z$ pole observables are 
summarized in Table~\ref{zpole} and the non-$Z$ pole observables in 
Table~\ref{nonzpole}. For more details and further references we refer to our 
recent reviews~\cite{Erler98}.

The theoretical evaluation uses the FORTRAN package GAPP~\cite{Erler00} 
dedicated to the Global Analysis of Particle Properties. GAPP attempts to 
gather all available theoretical and experimental information from precision 
measurements in particle physics. It treats all relevant SM inputs and new
physics parameters as global fit parameters. For clarity and to minimize CPU 
costs it avoids numerical integrations throughout. GAPP is based on the  
$\overline{\rm MS}$ renormalization scheme which demonstrably avoids large 
expansion coefficients. 

\begin{table}[thb]
\centering
\begin{tabular}{|l|l|c|c|c|c|c|c|c|}
\hline
\multicolumn{2}{|l|}{}
&$Z_\chi$&$Z_\psi$&$Z_\eta$&$Z_{LR}$&$Z_{SM}$&$Z_{\rm string}$& SM \\
\hline\hline
&& 551(591) & 151(162) & 379(433) & 570(609) & 822(924) & 582(618) & \\ 
\cline{2-8} 
&$\sin\theta$   &$-0.0006$& +0.0004 &$-0.0010$& +0.0002 &$-0.0015$&$-0.0002$&\\
                & $\sin\theta_{\rm min}$
                &$-0.0022$&$-0.0015$&$-0.0058$&$-0.0010$&$-0.0040$&$-0.0011$&\\
                & $\sin\theta_{\rm max}$
                & +0.0020 & +0.0021 & +0.0019 & +0.0022 & +0.0008 & +0.0008& \\
\cline{2-9} \raisebox{1.5ex}[0pt]{$\rho_0$ free}
&$\rho_0$          &0.9993 & 0.9974 & 0.9979 & 0.9995 &0.9982 &0.9996 &0.9996\\
&$\rho_0^{\rm min}$&0.9931 & 0.9923 & 0.9931 & 0.9917 &0.9933 &0.9986 &0.9985\\
&$\rho_0^{\rm max}$&1.0010 & 1.0017 & 1.0017 & 1.0013 &1.0018 &1.0011 &1.0017\\
\cline{2-9} 
&$\chi^2_{\rm min}$& 27.62 & 27.52 & 27.34 & 27.71 & 26.83 & 27.34 & 28.37 \\
\hline\hline
&& 545(582) & 146(155) & 365(408) & 564(602) & 809(894) & 578(612) & \\
\cline{2-8}        
& $\sin\theta$  &$-0.0003$& +0.0005 &$-0.0026$& +0.0003 &$-0.0019$&$-0.0002$&\\
$\rho_0=1$      & $\sin\theta_{\rm min}$
                &$-0.0020$&$-0.0013$&$-0.0062$&$-0.0009$&$-0.0041$&$-0.0011$&\\
                & $\sin\theta_{\rm max}$
                & +0.0015 & +0.0024 & +0.0011 & +0.0017 & +0.0003 & +0.0007 &\\
\cline{2-9}
&$\chi^2_{\rm min}$& 28.43 & 28.09 & 28.17 & 28.22 & 27.43 & 27.82 & 28.79 \\
\hline
\end{tabular}
\caption{Mass limits [in GeV] on extra $Z^\prime$ bosons and constraints
on $Z$--$Z^\prime$ mixing for two classes of Higgs sectors.
The upper part of the Table allows $\rho_0$ as a free fit parameter and 
corresponds to a completely arbitrary Higgs sector. The lower part assumes
$\rho_0 = 1$, but is arbitrary otherwise. The first (second) numbers 
correspond to the 95 (90)\% CL lower mass limits. Below this we show the 
central values  and the 95\% lower and upper limits on $\sin\theta$. Also 
shown are the central values and 95\% limits for $\rho_0$ as a fit parameter. 
Finally we indicate the minimal $\chi^2$ for each model. The last column
is included for comparison with the standard case of only one $Z$ boson. 
All results assume $M_Z \leq M_H \leq 1$ TeV.}
\label{arbit}
\end{table}

\begin{figure}[t]
\scalebox{0.5}{\includegraphics{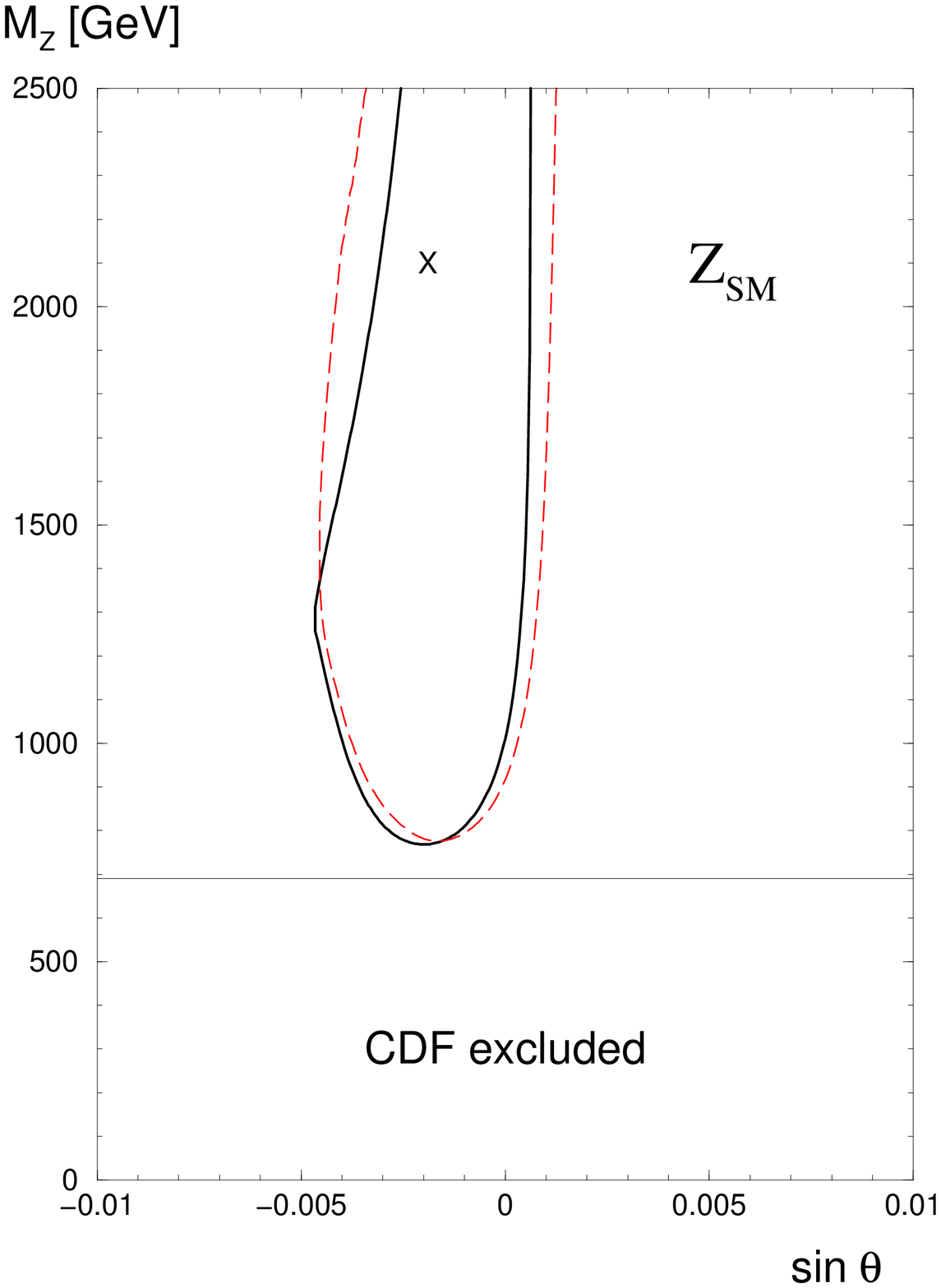}
               \includegraphics{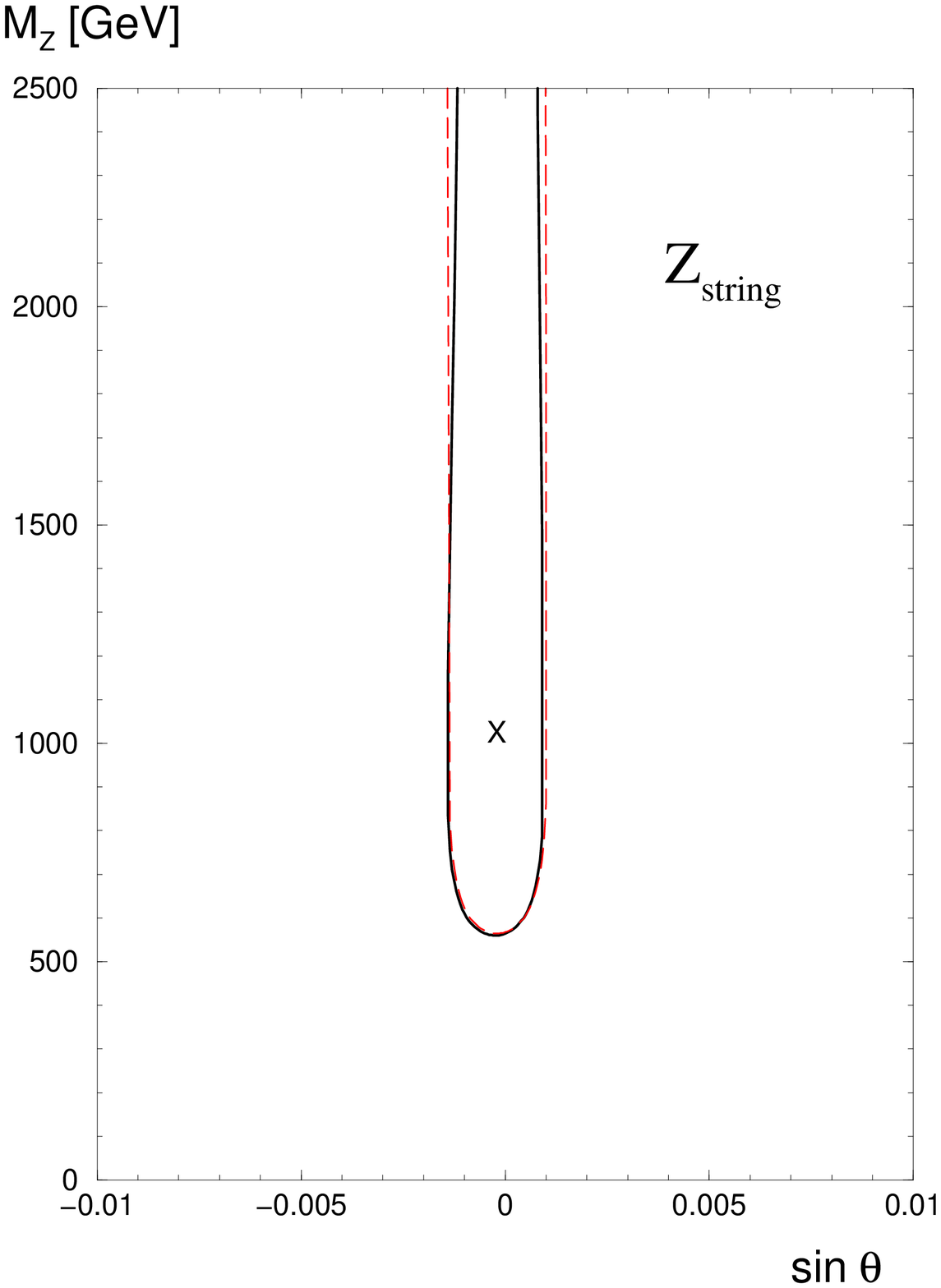}}
\caption{90\% CL contours for various $Z^\prime$ models. The solid contour
lines use the constraint $\rho_0 = 1$ (the cross denotes the best fit location
for the $\rho_0 = 1$ case), while the long-dashed lines are for arbitrary 
Higgs sectors. Also shown are the additional constraints in minimal Higgs 
scenarios for several VEV ratios as discussed in the text. The lower direct 
production limits from CDF~\cite{Abe97} are also shown. They assume that all
exotic decay channels are closed, and have to be relaxed by about 100 to 150 
GeV when all exotic decays (including channels involving superparticles) are 
kinematically allowed~\cite{Abe97}.}
\label{fig}
\end{figure}

In Tables~\ref{arbit} and~\ref{specific} and Figure~\ref{fig} we present our
main results. We list lower limits on $Z^\prime$ boson masses for a variety of 
cases. Note, that the new physics, i.e., the $Z^\prime$s and the extra Higgs 
bosons, decouple and that the SM ($M_{Z^\prime} = \infty$, $\theta = 0$) is 
well within the allowed regions of Figure~\ref{fig}. As a consequence a 
rigorous Bayesian integration over the $M_{Z^\prime}$ probability distribution 
diverges\footnote{The Bayesian confidence integral in Eq.~(4.13) of 
reference~\cite{Cho98} is not well-defined unless a non-trivial Jacobian
is implicitly included.}. Therefore, we approximate the 95 (90)\% CL limits by 
requiring $\Delta \chi^2 = 3.84$ (2.71), which is motivated by a reference 
univariate normal distribution. Similarly, we define the 90\% allowed region in
the $(M_{Z^\prime},\sin \theta)$-plane by $\Delta \chi^2 < 4.61$, here 
referring to a bivariate Gaussian. 

\begin{figure}[p]
\scalebox{0.5}{\includegraphics{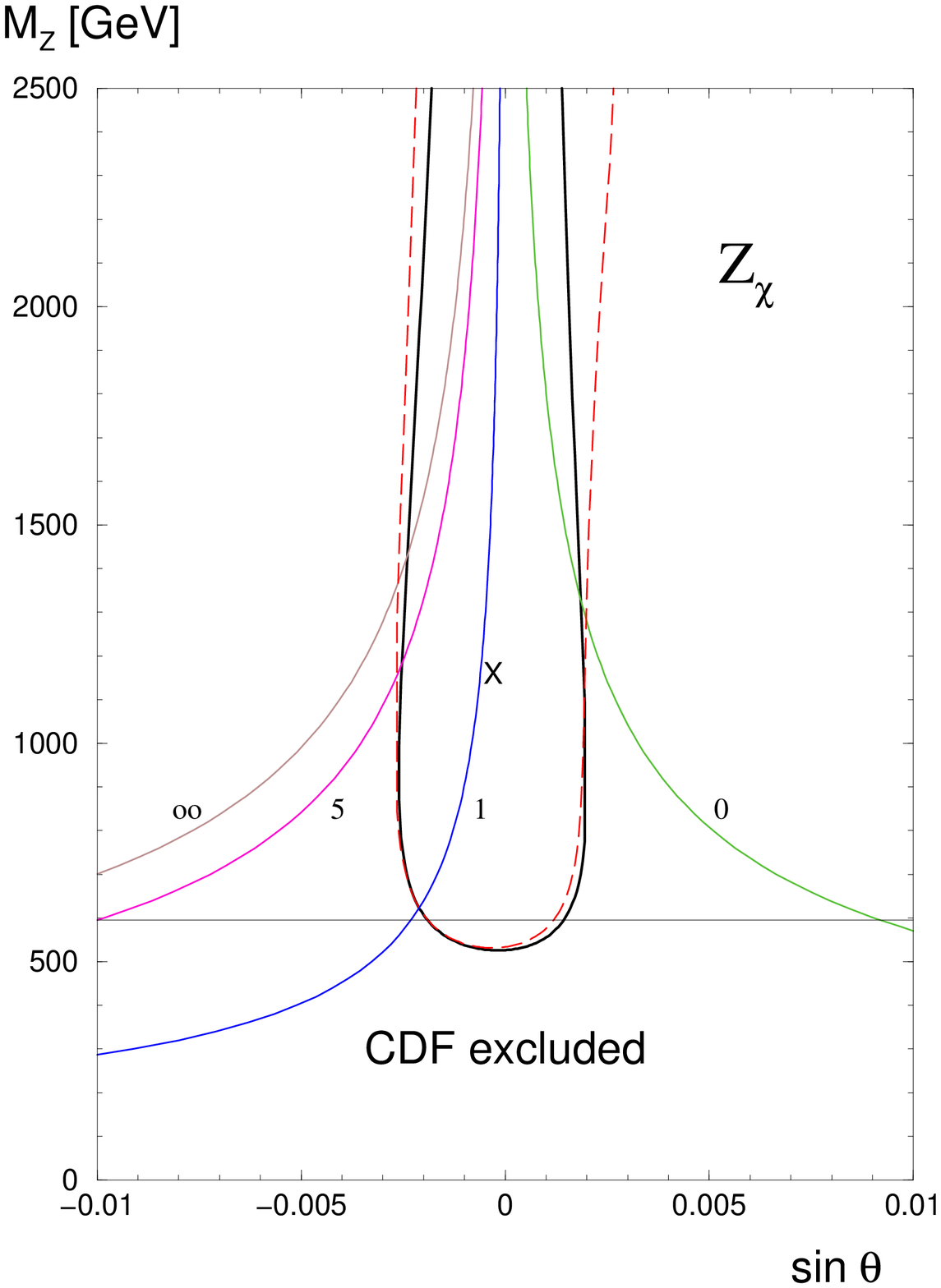}
               \includegraphics{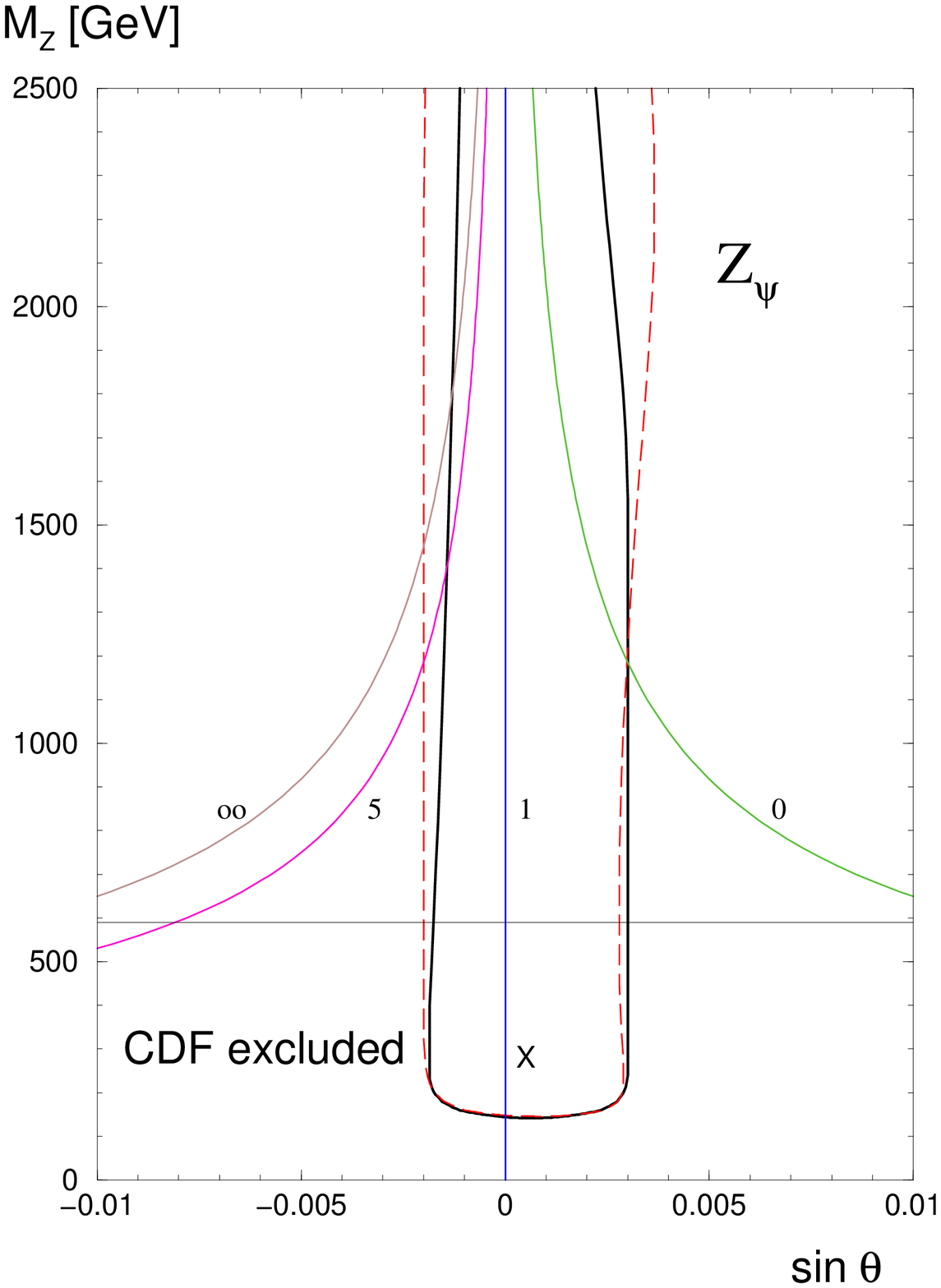}}
\scalebox{0.5}{\includegraphics{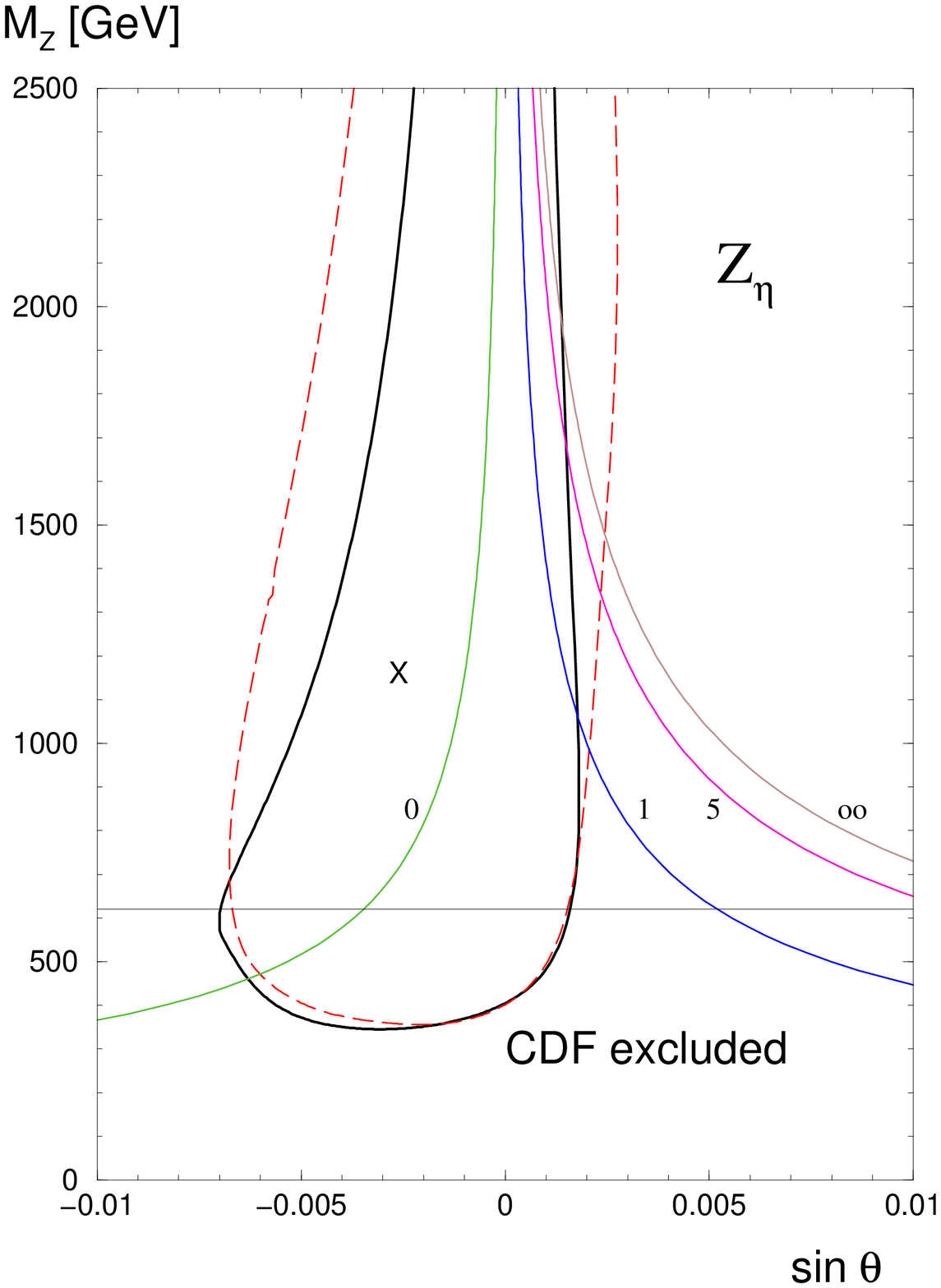}
               \includegraphics{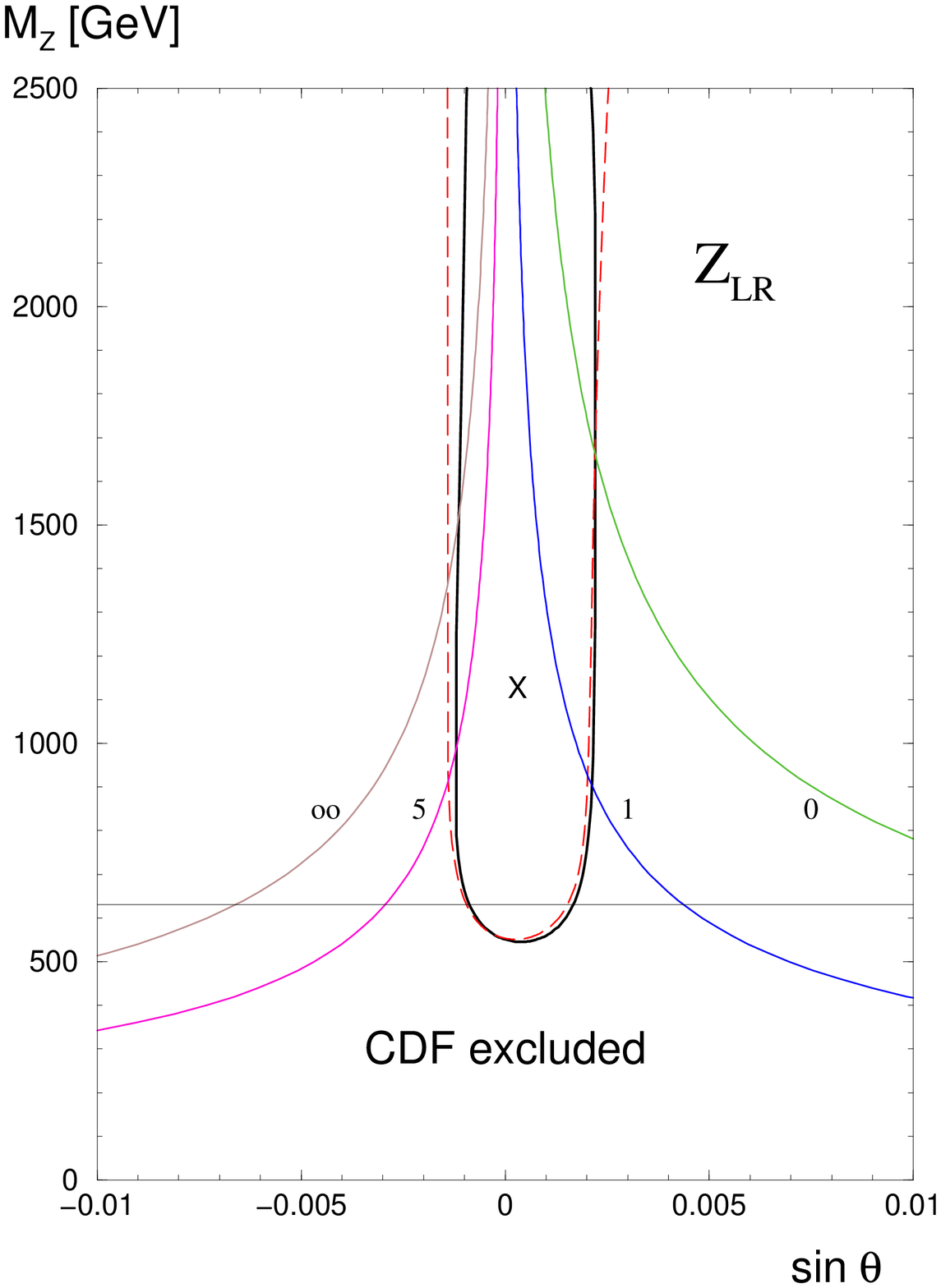}}
\end{figure}

The first set of fits in Table~\ref{arbit} is for the most general case of 
completely arbitrary Higgs sector, while the second set is for $\rho_0 = 1$. 
Below the mass limits we show the best fit values and 95\% lower and upper 
limits on the mixing parameter $\sin\theta$. Results on $\rho_0$ and the 
$\chi^2$ minimum are also shown. For comparison we have included the 
$SU(2)\times U(1)$ case in the last column. 

We note that the $Z_{LR}$ boson is equivalent to the $Z_\chi$ boson with a 
non-vanishing kinetic mixing term,
$$ - {\sin\chi\over 2} B^{\mu\nu} Z_{\mu\nu}^\prime . $$ 
It can be absorbed by a shift of the gauge couplings proportional to the third
component of the right-handed isospin generator and a rescaling of the coupling
ratio $\lambda$. In models with given Higgs structure the parameter 
$C \rightarrow C - \sqrt{3\over 5\lambda_g} \sin\chi$ is shifted, as well,
while $\rho_0$ is unaffected. The limits on masses and mixings of the $Z_\chi$ 
and $Z_{LR}$ bosons, shown in Tables~\ref{arbit} and~\ref{specific} (with
$\lambda = 1$), are indeed quite similar.

Our mass limits on extra $Z^\prime$ bosons are somewhat stronger than those 
from a recent analysis~\cite{Cho98} of $Z^\prime$s in supersymmetric $E_6$ 
models~\cite{Hewett89}. The differences are due to a slightly different
and more recent data set in our analysis, different implementations of 
radiative corrections, different statistical methods\footnote{The difference
goes beyond the more common choices of Bayesian versus frequentist kind of
approaches. The authors of reference~\cite{Cho98} choose to allow three
fit parameters, of which only two are independent. This is a problematic
procedure when parameter estimation is desired and renders confidence
intervals ambiguous.}, and our alternative 
evaluation of the photonic vacuum polarization effects~\cite{Erler99} with 
advantages for global fits. Moreover, reference~\cite{Cho98} assumes the SUSY 
inspired range for the Higgs mass, $M_H < 150$~GeV~\cite{Carena96}. 

\begin{table}[h]
\centering
\begin{tabular}{|l|c|c|c|c|c|c|}
\hline
&$Z_\chi$ & $Z_\psi$ & $Z_\eta$ & $Z_{LR}$ \\
\hline\hline
$\sigma=0$ & 1368(1528) & 1181(1275) & \hspace{1ex}470(\hspace{1ex}498) & 
             1673(1799) \\
$\sigma=1$ & \hspace{1ex}643(\hspace{1ex}688) & 
\hspace{1ex}146(\hspace{1ex}156) & 1075(1235) & 
\hspace{1ex}925(\hspace{1ex}987) \\
$\sigma=5$ & 1210(1314) & 1393(1581) & 1701(1948) & \hspace{1ex}980(1076) \\
$\sigma=\infty$ & 1464(1601) & 1810(2039) & 1985(2277) & 1537(1711) \\
\hline
\end{tabular}
\caption{95 (90)\% CL lower mass limits on specific $Z^\prime$ bosons
as they appear in models of unification. Assumed are minimal Higgs structures
and $\rho_0 = 1$. Note that $\sigma$ is defined differently for the $Z_\psi$ 
and $Z_\eta$ models, and the $Z_\chi$ and $Z_{LR}$ models, respectively, 
as explained in the text. In particular, the versions of the $Z_\chi$ and 
$Z_{LR}$ models most often considered correspond to $\sigma = 0$.}
\label{specific}
\end{table}

Table~\ref{specific} lists results for specific Higgs charge assignments as 
they occur in ``minimal'' models:
For the $Z_\psi$ and $Z_\eta$ bosons, we assume an $SU(2)$ Higgs singlet
with a large VEV $s$ to ensure $M_{Z^\prime} \gg M_Z$, and in addition a pair 
of Higgs doublets with quantum numbers as in the ${\bf 5} + \bar{\bf 5}$ of
$SU(5)$ appearing in the {\bf 27} of $E_6$. They receive VEVs $v$ and
$\bar{v}$, where the combination $v^2 + \bar{v}^2$ is fixed by the measured
value of the Fermi constant. Thus, these models are described by an extra 
parameter $\sigma = |\bar{v}/v|^2$ which is analogous to $\tan^2 \beta$ in
supersymmetric extensions of the standard model. 

The $Z_\chi$ model does not depend on the ratio $|\bar{v}/v|$ so that
$C = 2/\sqrt{10}$ is predicted. If we add another Higgs doublet with quantum 
numbers like the SM leptons and VEV $x$, we have the extra parameter 
$\sigma = |x|^2/(v^2 + \bar{v}^2)$, and then $C \in 1/\sqrt{10}\, [-3,2]$. 
However, in those models of SUSY in which this Higgs doublet is identified 
with the superpartner of a SM lepton doublet, one has to require 
$x = \sigma = 0$ to avoid severe problems with charged-current universality. 

The Higgs content of the $Z_\chi$ model can be lifted to an appropriate 
Higgs structure for a $Z_{LR}$ model (LR 1), transforming under 
$SU(2)_L \times SU(2)_R \times U(1)_{B-L}$ as 
$({\bf 2},\bar{\bf 2},0) + ({\bf 2},{\bf 1},1/2) + ({\bf 1},{\bf 2},1/2)$.
The same definitions and remarks apply, except that here 
$C \in \sqrt{3/5}\, [-1/\alpha,\alpha]$. Another possibility (LR 2) is a Higgs 
sector transforming as
$({\bf 2},\bar{\bf 2},0) + ({\bf 3},{\bf 1},1) + ({\bf 1},{\bf 3},1)$ which
results in $C = \sqrt{3/5}\, \alpha$. There are no $SU(2)$ triplets in a 
${\bf 27}$ of $E_6$, but one might find them in string models without an 
intermediate GUT group, if they are realized at higher Kac-Moody levels 
($k > 1$). In the context of SUSY the Higgs fields carrying $B-L$ charge 
have to be supplemented by extra fields with opposite charge to cancel 
anomalies and (in the case of LR~1) to generate fermion masses.

There is no analog of a minimal Higgs sector for the sequential $Z_{SM}$. In 
the $Z_{\rm string}$ model $\rho_0 = 1$ is predicted, and even the $Z^\prime$ 
mass and the $Z$--$Z^\prime$ mixing can be calculated in principle. Universal 
high scale boundary conditions yield too large a value for $\theta$ and are 
excluded~\cite{Langacker98}. Non-universal boundary conditions on the soft
SUSY breaking terms can yield acceptable mixing. In any case, the concrete 
realization of the soft terms depends strongly on the SUSY breaking and 
mediation mechanisms. We parametrize our lack of understanding by allowing
an arbitrary Higgs sector (except for $\rho_0 = 1$).

In conclusion, indirect constraints from high precision observables on and off 
the $Z$ pole continue to play an important role for searches for new physics,
and in particular for extra gauge bosons. The obtained mass limits are 
competitive with current direct searches at colliders with the highest 
attainable energies. Moreover, no assumptions about the absence of exotic decay
channels are necessary\footnote{However, we ignored contributions from
possible exotic states to the $S$ parameter. The potentially much larger 
contributions to $T$ are accounted for in our fits with $\rho_0$ allowed.}. 
The indirect constraints are even much stronger in specific models with known 
Higgs structure, where lower limits are typically in the 1~to~2~TeV range. 
Finally, $Z$--$Z^\prime$ mixing effects are severely constrained 
to the sub per cent level in all cases. 

\section*{Acknowledgement:}
This work was supported in part by U.S. Department of Energy Grant 
EY--76--02--3071.

\end{document}